\documentclass[11pt,twoside]{article}

\usepackage{asp2006}
\usepackage{epsf}
\usepackage{psfig}
\usepackage{lscape}

\begin{document}
\title{A Simple Isolation Criterion based on 3D Redshift Space \boldmath $(\alpha,\delta,z)$ Mapping}  
\author{Oded Spector and Noah Brosch}   
\affil{Wise Observatory and the Beverly and Raymond Sackler School of Physics and Astronomy,
Tel Aviv University, Tel Aviv 69978, Israel}    

\begin{abstract} 
We selected a sample of galaxies, extremely isolated in 3D redshift space, based on data from NED and the ongoing ALFALFA HI (21cm) survey. A simple selection criterion was employed: having no neighbors closer than 300 km$\cdot$s$^{-1}$ in 3D redshift space $(\alpha,\delta,z)$. The environments of galaxies, selected using this criterion and NED\footnote{The NASA/IPAC Extragalactic Database (NED) is operated by the Jet Propulsion Laboratory, California Institute of Technology, under contract with the National Aeronautics and Space Administration.} data alone, were analyzed theoretically using a constrained simulation of the local Universe, and were found to be an order of magnitude less dense than environments around randomly selected galaxies. One third of the galaxies selected using NED data alone did not pass the criterion when tested with ALFALFA data, implying that the use of unbiased HI data significantly improves the quality of the sample.
\end{abstract}


\section{Introduction}
The research described in this paper is part of an extensive study of star formation and evolution of galaxies in different environments and of various morphological types, conducted by our group for the past few decades \citep{1983PhDT.........1B, 1998ApJ...504..720B, 2008MNRAS.390..408Z}. Specifically, we are studying galaxies in the most extremely underdense regions of the local Universe. These galaxies are particularly interesting, since they have evolved with little or no environmental interference, and therefore are good test beds for calibrating galaxy evolution models. Furthermore, when compared to galaxies in denser regions they illuminate the overall effects of the environment on the evolution of galaxies.

The first and possibly most crucial step in this study is selecting a sample of extremely underdense region galaxies, referred to here as Isolated Galaxies (IGs). By definition these IGs are very rare, and therefore at moderate redshifts the sample is fairly small, including only a few dozen galaxies. Here we describe the method used for selecting the sample and the validation process used to ensure that these are indeed galaxies in the most 
extremely underdense regions of the local Universe.

The sample is currently being imaged in optical wide bands and $H_\alpha$ to measure the current star formation and the evolutionary histories of these galaxies. The observations are carried out at the Wise Observatory. The ALFALFA blind HI survey \citep{2005AJ....130.2598G, 2009ApJS..183..214M}, which covers 7000 deg$^2$ of the sky to redshift 17000 km$\cdot$s$^{-1}$ with low detection limit and excellent positioning accuracy ($\sim$0.1$\arcmin$), is a key data source for selecting the sample and analyzing its HI content. Further observational data are obtained from NED and SDSS \citep{2009ApJS..182..543A}.

\section{The Isolation Criterion}

In the last few decades great advances were made in redshift surveys, which now map the local Universe in redshift space with great precision. Before these were available isolated galaxies had to be identified using projected coordinates alone $(\alpha,\delta)$, i.e.~searching in two-dimensional space (2D). Sophisticated algorithms were used to estimate radial distance, e.g.~via the angular sizes of galaxies, such as done for the classical \cite{1973SoSAO...8....3K} catalog of isolated galaxies. Nowadays, when the local Universe is mapped in fine detail, it is possible to perform three-dimensional (3D) searches. The advantages of using such strategy are simplicity and straightforwardness, not having to assume anything about the characteristics of the galaxies (size, magnitude, etc.).

However, using redshift mapping introduces two difficulties. First is the incompleteness of most redshift databases. A galaxy that seems to be isolated might have neighbors for which a redshift was never measured. Second is the error in radial distance, introduced by peculiar velocities. Using redshift data one performs a search in 3D redshift space, the mathematical representation of the projected $(\alpha,\delta)$ and radial (redshift) coordinates. It should be kept in mind that mapping in 3D redshift space $(\alpha,\delta,z)$ can differ significantly from the true mapping in position space $(\alpha,\delta,r)$. For example, although close in position space, two galaxies in a cluster might have very different redshifts due to cluster velocity dispersion and will therefore seem distant in 3D redshift space.

In this work we have chosen to use a simple selection criterion. A galaxy is considered an IG and is included in the sample if it has no known neighbors closer than 300 km$\cdot$s$^{-1}$ in 3D redshift space, and its redshift is in the range $2000<cz<7000$ km$\cdot$s$^{-1}$.

The redshift range is limited to 7000 km$\cdot$s$^{-1}$ to have better completeness of redshift data around each galaxy. The reason for the lower limit of 2000 km$\cdot$s$^{-1}$ is to keep the sky area that has to be searched around each galaxy, relatively low (at 2000 km$\cdot$s$^{-1}$ neighbors have to be searched for as far as a projected distance of 8.6$\deg$).

\section{The Sample}
The search criterion was applied to two sky regions, one in the spring sky: 7h30m$<\alpha<$16h30m, $4\deg<\delta<16\deg$, $2000<cz<7000$ km$\cdot$s$^{-1}$, and the other in the autumn sky: 22h00m$<\alpha<$03h00m, $24\deg<\delta<28\deg$, $2000<cz<7000$ km$\cdot$s$^{-1}$. Both regions were already covered by AFLAFA. The ``spring'' sky region is also fully covered by SDSS DR7 \citep{2009ApJS..182..543A}.

IGs were searched for in these regions first based only on NED data of January 2009. In the ``spring'' sky region 27 IGs were found out of 2826 candidates ($\sim$1.0\%). In the ``autumn'' sky region 6 IGs were found out of 244 candidates ($\sim$2\%). ALFALFA data were then used to find possible neighbors to the ``spring'' IGs. Out of 27 IGs, nine (a third) had ALFALFA neighbors. Four galaxies had more than one ALFALFA neighbor, and one galaxy had 7 ALFALFA neighbors, none of which had redshift data from optical spectra!

The search was repeated for the ``spring'' sky region in a different fashion. The search criterion was first applied to ALFALFA’s data, and then the isolation of the resulting galaxies was tested using NED. Five additional IGs were found this way (none of which had redshift data in NED).

Figure 1 compares the number density $n$ around a typical IG from the sample (UGC 12123) to the number density around a typical field galaxy (LEDA 166859) and around M87, located at the center of the Virgo cluster. For each of the three galaxies $n$ is shown as a function of $r$, the radius of a sphere around the galaxy for which $n$ was calculated.

UGC 12123 has two neighbors within 5 Mpc$\cdot$h$^{-1}$: one 3.06 Mpc$\cdot$h$^{-1}$ away (detected by ALFALFA) and the other 4.35 Mpc$\cdot$h$^{-1}$ away. As can be seen in Figure 1, the neighborhood density of the IG (UGC 12123) is about one order of magnitude lower than that of the typical field galaxy (LEDA 166859) and more than two orders of magnitude lower than that of the cluster galaxy (M87).

\section{Validation}

In order to validate that when applying the search criterion to NED data, one obtains a sample of galaxies in extremely underdense regions of the Universe, an independent dataset was used. This dataset was the Box 160 constrained simulation of the local Universe (Hoffman \& Gottl\"{o}ber 2008, private communication; Gottl\"{o}ber S. \& Klypin 2008; Forero-Romero et al.~2009). Box 160 is a $\Lambda$CDM (WMAP3) based mock universe that emulates large structures in the local Universe (Virgo, Coma, Local Supercluster, etc.)~with box side length of 160 Mpc$\cdot$h$^{-1}$ and $1024^{3}$ particles, 2.54$\cdot$10$^{8}$ $M_{\sun}\cdot$h$^{-1}$ each.

The Box 160 dataset was used to create mock ``observable'' data, consisting of $(\alpha,\delta)$ coordinates and observable magnitudes for all galaxies. Redshift data were added to the mock ``observable'' data for 38\% of the galaxies with observable magnitude $m\leq18.5$ (randomly selected) and non of the galaxies with $m>18.5$. This was done to simulate the dilution of redshift data in the NED dataset used for selecting the sample from the ``spring'' sky region.

A ``mock sample'' was created by applying the search criterion to the ``mock observable'' data. The ``true neighborhood'' of the ``mock sample'' IGs was analyzed using the full Box 160 data, i.e. using ``true'' coordinates of all the neighbors including those which the ``mock observable'' dataset missed.

The results are shown in Figure 2, where the likelihood function of the number density around a galaxy is plotted for the ``mock sample'' and for all the Box 160 galaxies (``all''). As can be seen, the IGs selected by this method (based on data that simulate NED) are located in significantly underdense neighborhoods compared to the general population.

As we have seen, the use of ALFALFA significantly improves the quality of the sample, i.e.~the likelihood function is expected to be shifted to lower densities. However, this could not be simulated because a mock universe dataset that includes HI mass data is not available at present. Without HI data it cannot be discerned which simulated objects would have been observable by ALFALFA.

\begin{figure}[!t]
  \plottwo{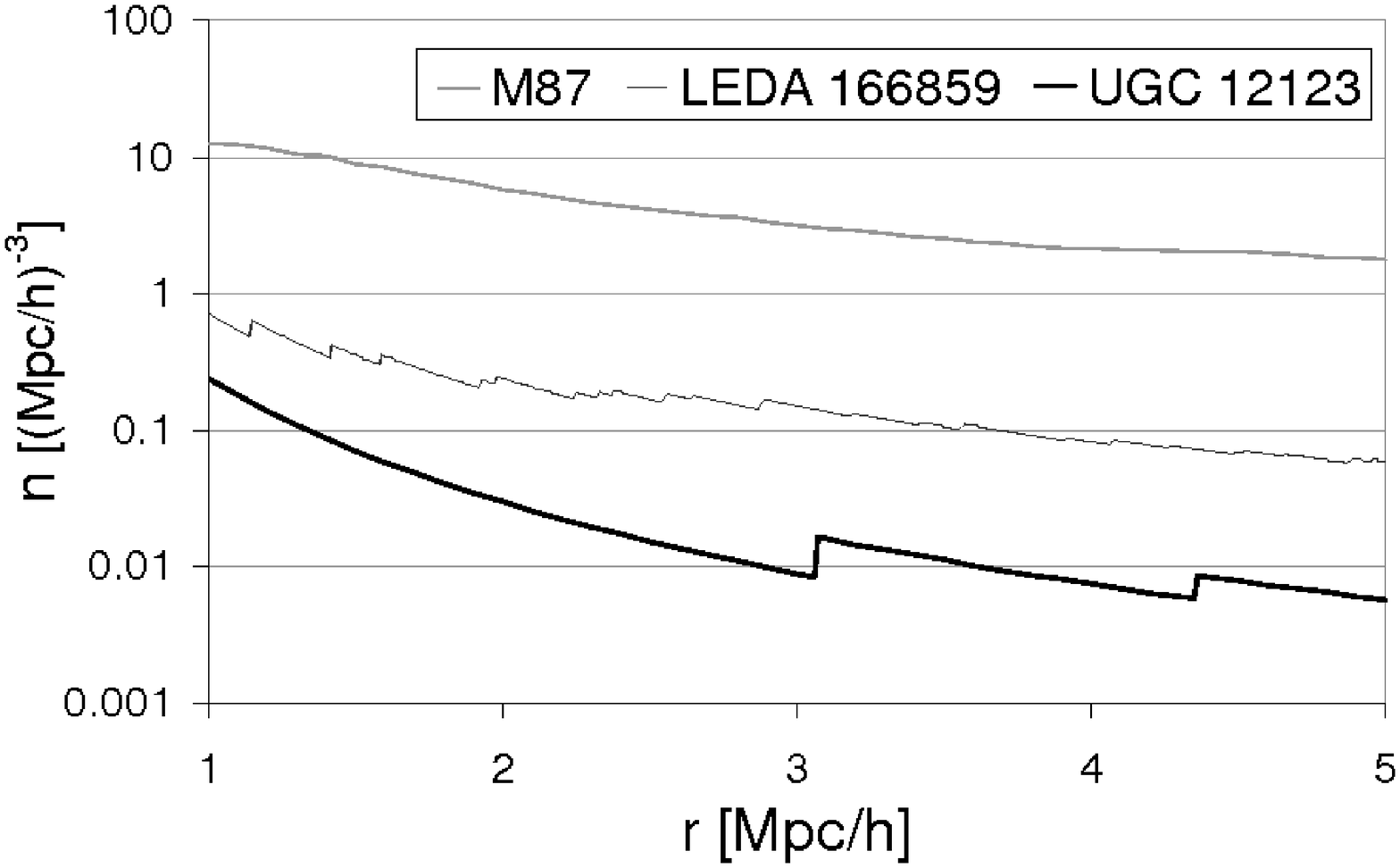}{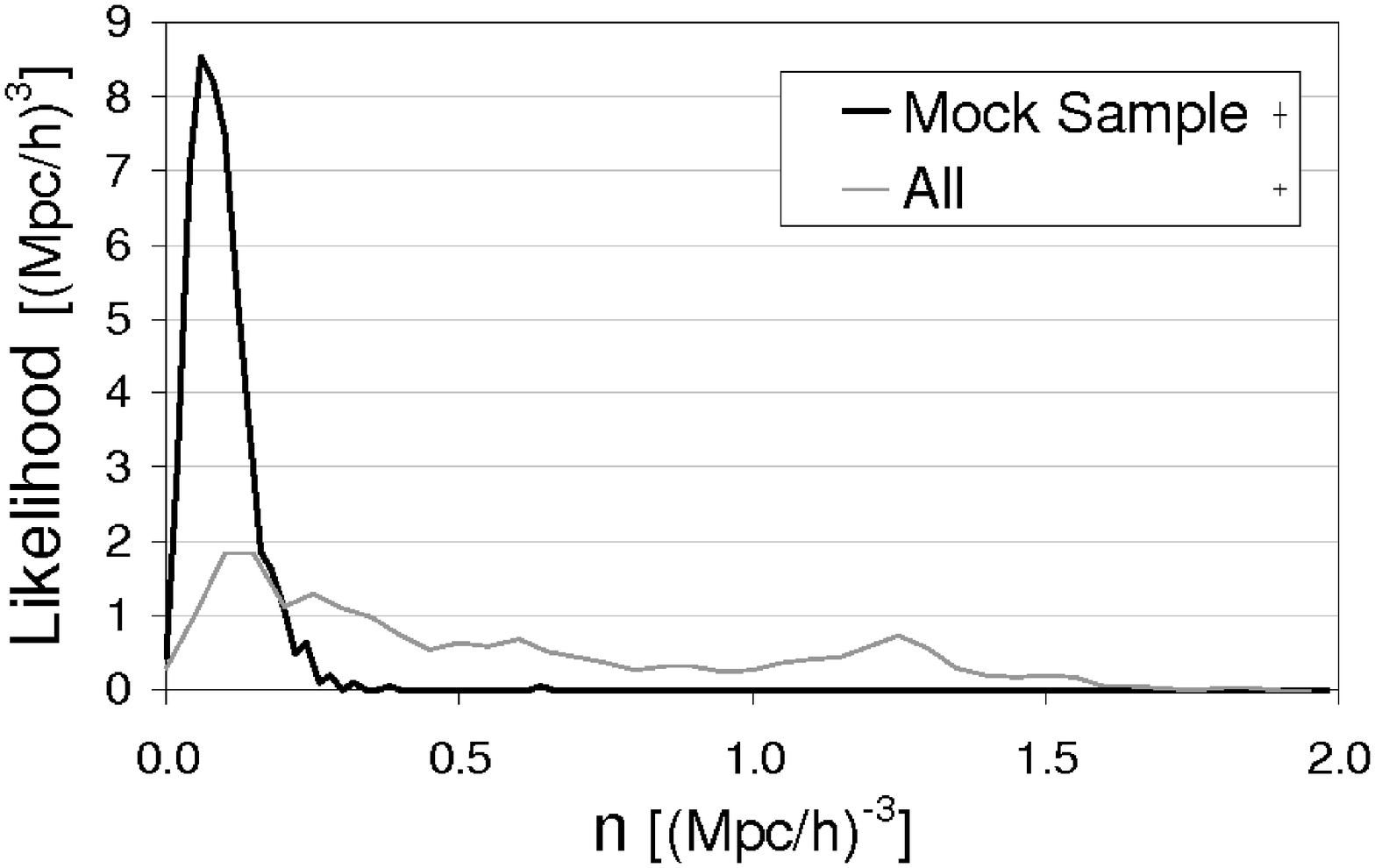}
  \caption{The number density $n$ in a sphere of varying radius $r$ around three galaxies:
UGC 12123 – an IG from the sample, LEDA 166859 – a field galaxy, M87 – the central galaxy of the Virgo cluster. \label{Fig1}}
  \caption{ Likelihood function of the number density $n$ in a 5 Mpc$\cdot$h$^{-1}$ radius sphere around galaxies of the ``mock sample'', and around galaxies of the entire mock universe (All). Errors are indicated by the + signs in the legend. \label{Fig2}}
\end{figure}

\section{Conclusions}

\begin{itemize}
  \item Redshift surveys provide enough information for choosing IG samples based on 3D mapping.

  \item The simple isolation criterion (having no known neighbors within 3 Mpc$\cdot$h$^{-1}$), when applied to redshifts in the range $2000<cz<7000$ km$\cdot$s$^{-1}$, provides a sample of galaxies in the low density regions of the local Universe.

  \item ALFALFA data are extremely useful in eliminating false positives from the sample due to the redshift data it provides for low-luminosity galaxies with high HI masses.

\end{itemize}

\acknowledgements We are grateful to Martha Haynes, Riccardo Giovanelli, and the entire ALFALFA team for providing an unequalled HI data set. We are grateful to Yehuda Hoffman, Stefan Gottl\"{o}ber and Ofer Metuki for providing the Box 160 simulation dataset. This research has made use of the NASA/IPAC Extragalactic Database (NED) which is operated by the Jet Propulsion Laboratory, California Institute of Technology, under contract with the National Aeronautics and Space Administration.


\begin{thebibliography}{}


\bibitem[Abazajian et al.(2009)]{2009ApJS..182..543A} Abazajian, K.~N., et 
al.\ 2009, \apjs, 182, 543 

\bibitem[Brosch(1983)]{1983PhDT.........1B} Brosch, N.\ 1983, Ph.D.~Thesis, 

\bibitem[Brosch et al.(1998)]{1998ApJ...504..720B} Brosch, N., Heller, A., 
\& Almoznino, E.\ 1998, \apj, 504, 720 

\bibitem[Forero-Romero et al.(2009)]{2009MNRAS.396.1815F} Forero-Romero, 
J.~E. et al.\ 2009, \mnras, 396, 1815 

\bibitem[Giovanelli et al.(2005)]{2005AJ....130.2598G} Giovanelli, R., et 
al.\ 2005, \aj, 130, 2598 

\bibitem[Gottl{\"o}ber 
\& Klypin(2008)]{2008arXiv0803.4343G} Gottl{\"o}ber, S., \& Klypin, A.\ 2008, arXiv:0803.4343 

\bibitem[Karachentseva(1973)]{1973SoSAO...8....3K} Karachentseva, V.~E.\ 
1973, Soobshcheniya Spetsial'noj Astrofizicheskoj Observa., 8, 3 

\bibitem[Martin et al.(2009)]{2009ApJS..183..214M} Martin, A.~M. et al.\ 2009, \apjs, 183, 214 

\bibitem[Zitrin 
\& Brosch(2008)]{2008MNRAS.390..408Z} Zitrin, A., \& Brosch, N.\ 2008, \mnras, 390, 408 

\end{thebibliography}
\end{document}